\documentclass[prl,twocolumn, showpacs,preprintnumbers,superscriptaddress]{revtex4}
\usepackage{amssymb}
\usepackage{graphicx}
\usepackage{natbib}
\usepackage{amsmath}
\begin{document}
\hyphenation{Studien-stiftung}

%**********Title of paper***********
\title{Mixing-induced anisotropic correlations in molecular crystalline systems}

%***********Authors************
\author{A. Aufderheide}\affiliation{Universit\"at T\"ubingen, Insitut f\"ur
  Angewandte Physik, Auf der Morgenstelle 10, 72076 T\"ubingen, Germany}

\author{K. Broch}\affiliation{Universit\"at T\"ubingen, Insitut f\"ur
  Angewandte Physik, Auf der Morgenstelle 10, 72076 T\"ubingen, Germany}

\author{J. Nov$\mathrm{\acute{a}}$k}\affiliation{Universit\"at T\"ubingen,
  Insitut f\"ur Angewandte Physik, Auf der Morgenstelle 10, 72076 T\"ubingen,
  Germany}

\author{A. Hinderhofer}\affiliation{Universit\"at T\"ubingen, Insitut f\"ur
  Angewandte Physik, Auf der Morgenstelle 10, 72076 T\"ubingen, Germany}

\author{R.Nervo}\affiliation{ESRF, 6 Rue Jules Horowitz, BP 220, 38043
  Grenoble Cedex 9, France}

\author{A. Gerlach}\affiliation{Universit\"at T\"ubingen, Insitut f\"ur
  Angewandte Physik, Auf der Morgenstelle 10, 72076 T\"ubingen, Germany}

\author{R. Banerjee}\affiliation{Universit\"at T\"ubingen, Insitut f\"ur
  Angewandte Physik, Auf der Morgenstelle 10, 72076 T\"ubingen, Germany}

\author{F. Schreiber}\email{frank.schreiber@uni-tuebingen.de}\affiliation{Universit\"at
  T\"ubingen, Insitut f\"ur Angewandte Physik, Auf der Morgenstelle 10, 72076
  T\"ubingen, Germany} 

\date{\today}

%%%%%%%%%%%%%%%%%%%%%%%%%%%%%%%%%%%%%%%%%%%%%%%%%%%%%%%%%%%%%%%%%%
%%%%%%%%%%%%%
%%%%%%%%%%%%%           ABSTRACT          %%%%%%%%%%%%%%%%%%%%%%%%
%%%%%%%%%%%%%
%%%%%%%%%%%%%%%%%%%%%%%%%%%%%%%%%%%%%%%%%%%%%%%%%%%%%%%%%%%%%%%%%%
\begin{abstract}
  We investigate the structure of mixed thin films composed of pentacene (PEN)
  and diindenoperylene (DIP) using X-ray reflectivity and grazing incidence
  X-ray diffraction. For equimolar mixtures we observe vanishing in-plane
  order coexisting with an excellent out-of-plane order, a yet unreported
  disordering behavior in binary mixtures of organic semiconductors, which are
  crystalline in their pure form. One approach to rationalize our findings is
  to introduce an anisotropic interaction parameter in the framework of a mean
  field model. By comparing the structural properties with those of other
  mixed systems, we discuss the effects of sterical compatibility and chemical
  composition on the mixing behavior, which adds to the general understanding
  of interactions in molecular mixtures.
\end{abstract}

\maketitle

%%%%%%%%%%%%%%%%%%%%%%%%%%%%%%%%%%%%%%%%%%%%%%%%%%%%%%%%%%%%%%%%%%
%%%%%%%%%%%%%
%%%%%%%%%%%%%           INTRODUCTION          %%%%%%%%%%%%%%%%%%%%
%%%%%%%%%%%%%
%%%%%%%%%%%%%%%%%%%%%%%%%%%%%%%%%%%%%%%%%%%%%%%%%%%%%%%%%%%%%%%%%%
Many modern materials and devices consist of rather complex mixtures. This is
also true for organics, which have multiple applications in optoelectronic
devices, such as organic photovoltaic systems with strongly increasing
interest in recent years \cite{Forrest2004N_428, Bruetting_2005,
  Brabec2001Afm_11}.  In binary systems, not only the nominal concentration of
two components A and B is relevant, but also the degree of intermixing, the
crystalline order, and the morphology, as well as the characteristic length
scales involved.  While these structural and morphological features have a
significant impact on the device performance~\cite{opitz_ieee10}, the
underlying driving forces for structure formation in molecular materials are
not well understood from a fundamental perspective.  Compared to mixtures of
elemental systems such as many binary alloys, for mixed organic systems
additional issues arise, such as the influence of steric
properties~\cite{Vogel_2010_JournalOfMaterialsChemistry_20,Hinderhofer_C_inprint}.

\begin{figure}
  \includegraphics[width=.9\columnwidth]{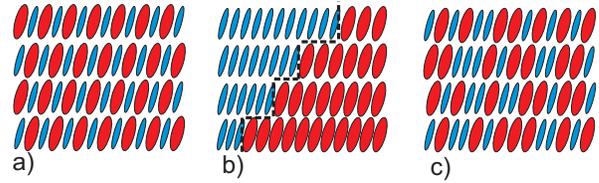}
  \caption{a)--c) Examples of possible growth scenarios for binary equimolar
    mixtures of molecules with a similar length but different width.}
  \label{fig:GrowthScen}
\end{figure}

\begin{figure}
  \includegraphics[width=.8\columnwidth]{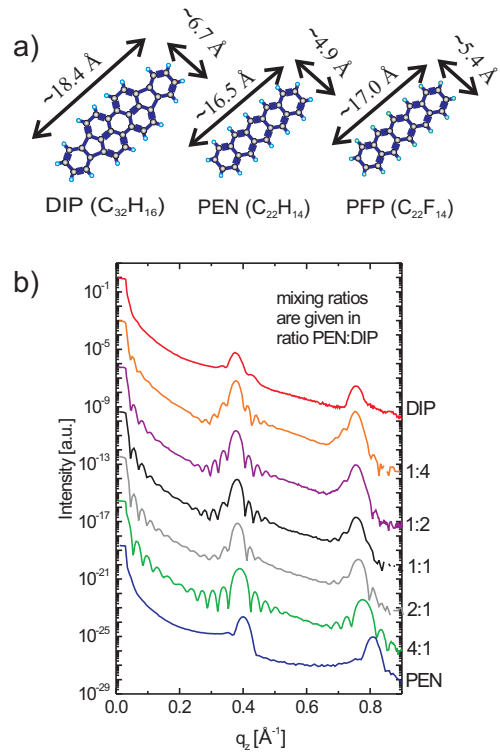}
  \caption{a) Chemical structure of DIP, PEN and PFP molecules b) XRR data for
    PEN:DIP films (offset for clarity).}
  \label{fig:XRR}
\end{figure}

A simple theoretical description of mixtures is provided by the `regular
solution model', which can also be applied to crystalline
systems~\cite{Kitaigorodsky1984}. Here, a binary mixture is described by a
mean-field approach with the free energy of mixing
\begin{equation}
  \Delta F_{mix}=k_B T [(x_A ~\mathrm{ln}~x_A+ x_B ~\mathrm{ln}~x_B)+\chi x_A x_B].
\label{eq:fmix}
\end{equation}
where $x_A$ and $x_B$ are the respective relative concentrations.  The $ln$
terms are due to entropy, which always favors mixing and the last term is
determined from the balance of the interaction energies with
\begin{equation}
\chi = \frac{Z}{k_B T}(W_{AA} + W_{BB} - 2W_{AB}),
\label{eq:chi}
\end{equation}
where $Z$ is the coordination number and $W_{AB}$ and $W_{AA}$ ($W_{BB}$) are
the interaction energies between dissimilar compounds A and B or between like
compounds A (B), respectively.  Generally, this leads to different mixing
scenarios~\cite{Kitaigorodsky1984}, depending on the value of $\chi$
(Fig. \ref{fig:GrowthScen}): \renewcommand{\labelenumi}{\alph{enumi})}
\begin{enumerate}
\item {$\chi < 0$: Intermixing; preference for A-B pairing}
\item {$\chi > 2$: Phase separation}
\item {$\chi \approx  0$: Random mixing determined by entropy}
\end{enumerate}
These scenarios have also been found for mixtures of organic semiconductors
(OSCs), such as pentacene (PEN, $\mathrm{C_{22}H_{14}}$), perfluoropentacene
(PFP, $\mathrm{C_{22}F_{14}}$) and diindenoperylene (DIP,
$\mathrm{C_{32}H_{16}}$), see Fig.~\ref{fig:XRR}a), although they have usually
not been discussed in terms of $\chi$~\cite{Hinderhofer2011J.Chem.Phys._134,
  Reinhardt_2012_JPCC_116_10917, Oteyza_2008_ThinSolidFilms_516, opitz_oe09,
  Chen_2011_Adv.Funct.Mater._21, Campione2009Chem.Mater._21,
  Salzmann_2008_JournalofAppliedPhysics_104,
  Vogel_2010_JournalOfMaterialsChemistry_20}. Importantly, in addition to
other known shortcomings of mean-field approaches, this model in its original
form does not take into account steric issues, anisotropies and predications
regarding crystallinity, although they may be incorporated. It is thus
\textit{a priori} not clear, if the scenario for mixtures of molecular
crystals is potentially richer than the three cases described above.

In this Letter, we report on {\it anisotropic} structure formation in thin
films of molecular mixtures of PEN and DIP (see Fig.~\ref{fig:XRR}a), which as
pure systems, exhibit excellent three-dimensional (3D) crystalline order. This
behavior changes dramatically upon mixing. Whereas along the surface normal
the mixed films exhibit nearly perfect order, for 1:1 blends the {\it
  in-plane} crystalline order essentially disappears, with some analogy to
(frozen) smectic order in liquid crystals.  We discuss these results in the
context of other recently studied binary mixtures of
OSCs~\cite{Hinderhofer2011J.Chem.Phys._134, Broch_2011_PRB_83_245307,
  Reinhardt_2012_JPCC_116_10917} and rationalize their ordering behavior by
proposing a model which includes steric properties and anisotropies and is
able to motivate not only \textit{mixing} but also ordering behavior in a
mixed system of organic semiconducting molecules.

%%%%%%%%%%%%%%%%%%%%%%%%%%%%%%%%%%%%%%%%%%%%%%%%%%%%%%%%%%%%%%%%%%
%%%%%%%%%%%%%
%%%%%%%%%%%%%           EXPERIMENTAL          %%%%%%%%%%%%%%%%%%%%
%%%%%%%%%%%%%
%%%%%%%%%%%%%%%%%%%%%%%%%%%%%%%%%%%%%%%%%%%%%%%%%%%%%%%%%%%%%%%%%%

Thin films containing PEN (purchased from Sigma Aldrich, 99.9$\%$ purity) and
DIP (purchased from Institut f$\mathrm{\ddot{u}}$r PAH Forschung Greifenberg,
Germany, 99.9$\%$ purity) were prepared by organic molecular beam deposition
(OMBD) on Si wafers covered with a native oxide layer similar to
Refs.~\onlinecite{Hinderhofer2011J.Chem.Phys._134, Broch_2011_PRB_83_245307}
at a base pressure of $ 2 \times 10^{-10}$~mbar. The substrate temperature was
kept constant at 26\,$^\circ$C. The films studied were grown with five
different mixing ratios of PEN:DIP (4:1, 2:1, 1:1, 1:2, 1:4), corrected for
the differences in the volumes of the unit cells and thus referring to molar
ratios. The estimated error of the stochiometry of the mixtures is about 10\%
determined by the error of the quartz-crystal-microbalance.

After growth, the samples were investigated by X-ray reflectivity (XRR) and
grazing incidence X-ray diffraction (GIXD) (for details see
Ref.~\onlinecite{Hinderhofer2011J.Chem.Phys._134}) at the ID10B beam-line of
the European Synchrotron Radiation Facility using a wavelength of
1.08\,$\mathrm{\AA}$ and a point detector with slits determining the
resolution. All measurements were performed under He-atmosphere to reduce air
scattering. Effects of air exposure and waiting time between film growth and
measurements on our results were excluded by additional real-time in situ
measurements in a vacuum chamber.

Figure~\ref{fig:XRR}b) shows XRR data for pristine PEN, DIP and their various
mixing ratios. All mixed films exhibit pronounced Laue and Kiessig
oscillations. The Laue oscillations result from a high out-of-plane
crystallinity of the sample with a coherence length similar to the total film
thickness of approximately 200~$\mathrm{\AA}$. The Kiessig fringes indicate
that the mixed films grow even more smoothly than the pure
ones. Interestingly, when varying the mixing ratio (PEN:DIP) from 1:4 to 4:1
we observed that the roughness decreases with increasing PEN ratio. A similar
behavior was observed of PEN:PFP
mixtures~\cite{Hinderhofer2011J.Chem.Phys._134}. The out-of-plane lattice
spacing shows a continuous shift to smaller values with increasing PEN ratio,
but with a non-linear dependence on the concentration (see supporting
material).
\begin{figure}
  \includegraphics[width=.8\columnwidth]{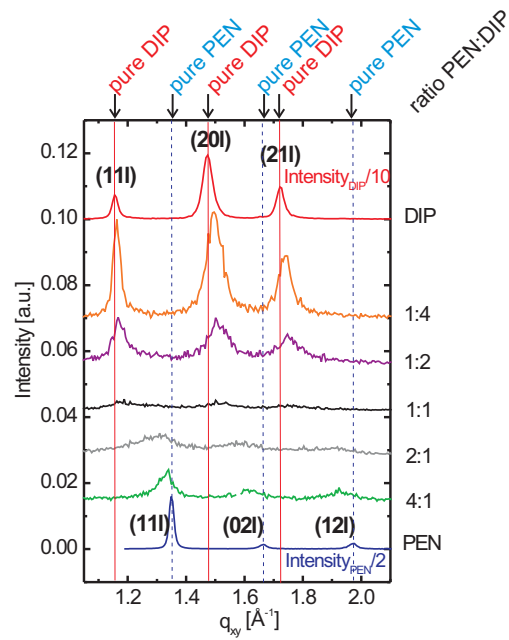}
  \caption{GIXD data obtained from PEN-DIP coevaporations with different
    mixing ratios. Data are offset for clarity. Note that the intensity of the
    DIP-peaks has been divided by a factor of 10 and the PEN-peak intensity by
    a factor of 2.}
  \label{fig:GIXD}
\end{figure}
Overall, the order is well-defined in the \textit{out-of-plane} direction, in
particular for the mixed films.

This is dramatically different for the \textit{in-plane} order. The in-plane
structure investigated by GIXD shows no order for the 1:1 mixing ratio (see
Fig.~\ref{fig:GIXD}). In particular, no peaks occur at $q_{xy}$ different from
the peak positions of the pure films. This is in contast to
PEN:PFP~\cite{Hinderhofer2011J.Chem.Phys._134} and
PFP:DIP~\cite{Reinhardt_2012_JPCC_116_10917} mixtures, for which new peaks
appear, which were assigned to a mixed crystal phase
(Fig.~\ref{fig:GrowthScen}a) with unit cells containing both compounds. The
small and broad features in the GIXD data of the equimolar mixture of PEN:DIP,
which occur in the region of the DIP peaks, are attributed to a small excess
of DIP molecules within the errorbar of the rate determination. GIXD data from
a different sample series (not shown here) do not even reveal traces of such
features, i.e. show a complete disappearance of the in-plane order for the
1:1-mixture.

For the non-equimolar mixtures, peaks appear at $q_{xy}$-positions in the
vicinity of those of the component dominating the mixture. The presence of
these peaks can be explained by minority molecules occupying sites in a
lattice formed by the more abundant molecular species. The resulting strain in
the lattice leads to the observed shift of the peak positions. The role of the
strain will be discussed in more detail below.

\begin{figure}
  \includegraphics[width=\columnwidth]{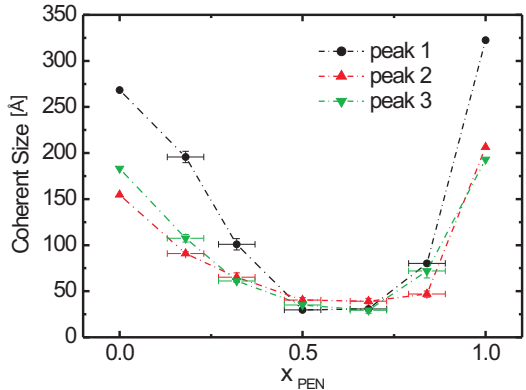}
  \caption{Lower limit for coherent in-plane size of crystallites versus
    relative molecular concentration of PEN ($x_{\mathrm{PEN}}$), derived from
    the FWHM of the in-plane peaks in Fig.~\ref{fig:GIXD}. Values of pure
    films and PEN1:DIP4 are resolution limited. The error bars of the coherent
    in-plane crystallite size are in the order of a few percent and thus in
    the range of the symbol size, whereas the error bar of the mixing ratio is
    10\% resulting from the inaccuracy of the QCM. The peaks are numbered in
    the order of ascending $q_{\mathrm{xy}}$ position.}
		\label{fig:Island}
\end{figure}

Figure~\ref{fig:Island} shows the in-plane coherent size of crystallites,
estimated from the GIXD peak widths using the Scherrer
formula~\cite{Warren_1991_DoverPubnIncDover}. The experimental resolution was
$\Delta q_{xy}\approx 0.01 ~\mathrm{\AA ^{-1}}$. Thus, except for the first
peak in the PEN1:DIP4 mixture and those of the pure films none of the GIXD
peaks observed for the mixtures are significantly broadened by the
resolution. Compared to the pure films the lower limit for the in-plane island
size is reduced by a factor of 2-10 and highly dependent on the mixing ratio.
Importantly, the reduced peak height for mixtures close to the equimolar
mixture is not simply due to peak broadening, but can be assigned to vanishing
in-plane order, while the out-of-plane order is preserved.

%%%%%%%%%%%%%%%%%%%%%%%%%%%%%%%%%%%%%%%%%%%%%%%%%%%%%%%%%%%%%%%%%%
%%%%%%%%%%%%%
%%%%%%%%%%%%%           DISCUSSION         %%%%%%%%%%%%%%%%%%%%%%%
%%%%%%%%%%%%%
%%%%%%%%%%%%%%%%%%%%%%%%%%%%%%%%%%%%%%%%%%%%%%%%%%%%%%%%%%%%%%%%%%
It is tempting to compare the ordering behavior of equimolar PEN:DIP mixtures
to that of liquid crystalline systems. Seen in this context, it would
correspond to a smectic C phase, characterized by crystalline order in one
direction (here the out-of-plane direction) and orientational order of tilted
molecules within the planes~\cite{Gennes_1993}, but no crystalline in-plane
order.  Indeed, changes in the ordering behavior upon mixing have also been
observed for liquid crystal
systems~\cite{Huang_2011_JournalofChemicalPhysics_134_124508,Rauch_2002_JCP_116_2213,Kapernaum_2010_C_11_2099}
but we emphasize that our system is conceptually different from liquid
crystalline systems, since the pure compounds show well-defined 3D-order over
a large temperature range.  To the best of our knowledge the anisotropic
change in the ordering behavior observed for equimolar mixtures of PEN:DIP is
a previously unreported effect for mixed systems of this class of molecular
compounds.

We attempt to rationalize this anomalous ordering behavior by extending the
mean field model (Eq.~\ref{eq:fmix}), within the limitations discussed in
Ref.~\onlinecite{Hinderhofer_C_inprint}, to mixtures of rod-like molecules
organized in layers. To do so we introduce an anisotropic interaction
parameter $\chi$ to take into account anisotropies in the intermolecular
interactions. $\chi$ splits into two components $\chi_{xy}$ and $\chi_{z}$,
which are defined according to Eq.~\ref{eq:chi}, for the in-plane and the
out-of-plane direction, respectively. Interactions arising from the chemical
composition, sterical properties (i.e.\ size, shape) and the average molecular
tilt angles $\theta_\mu$ ($\mu=[A,B]$) with respect to the surface normal of
the two compounds enter the nearest neighbour interactions energies $W_{ij}$
($i,j=[A,B]$).  In addition, strain and lattice deformation resulting from
differences in the sterical properties of two compounds, enter the free energy
$\Delta F$ of the system as a strain energy term $E^s$. For a layered system
$E^s$ has two components $E^{s}_{xy}$ and $E^{s}_{z}$, both depending on the
elastic constant tensor $\hat{C}$, the tilt angles $\theta_A$ and $\theta_B$
and the length ratio of the molecules $\beta_{\alpha}=l_{A\alpha}/l_{B\alpha}$
in the directions $\alpha=[xy,z]$. Here, $\beta_\alpha\approx 1$ and
$\beta_\alpha \gtrless 1$ indicate high and low steric compatibility,
respectively. Furthermore, due to the layer-wise growth using OMBD, and the
associated possibility of differences in the molecular concentration in the
different layers, we consider our films as a system of alternating layers.
Within these assumptions, the free energy per molecule $\Delta F$ describing
mixing and ordering of a mixed system can be written as:
\begin{align}
  \Delta F = & \frac{1}{2} k_\mathrm{B} T
	\{x_A\ln x_A+x_A^\star\ln x_A^\star 
  +x_B\ln x_B+x_B^\star\ln x_B^\star\nonumber\\
	&\;\;\;\;+\frac{1}{2}[\chi_{xy}(x_A x_B + x^\star_A x^\star_B)
	+\chi_z (x_A x^\star_B+
        x^\star_A x_B)]\}\nonumber\\
	&+E^{s}_{xy}(\hat{C},\beta_{xy},\theta_A,\theta_B)+E^{s}_z(\hat{C},\beta_z,\theta_A,\theta_B),
  \label{eq:freeE}
\end{align}
where $x_\mu$ and $x^\star_\mu$ stand for molar concentrations in alternate
layers along the vertical direction. They are related to the global molar
concentrations by $x^{g}_{\mu}=(x_{\mu}+x^{\star}_{\mu})/2$, with
$x^{g}_{A}+x^{g}_{B}=1$.  The interplay between three contributions to $\Delta
F$ predicts the ordering behavior of the system under mixing. The \textit{ln}
terms stem from the entropy, which favors statistical mixing. The interaction
terms contain the $\chi_{xy,z}$ and the strain terms, $E^{s}_{xy}$ and
$E^{s}_{z}$, are minimized by phase separation or the formation of a new
crystal structure.

\begin{figure}
  \includegraphics[width=.3\columnwidth]{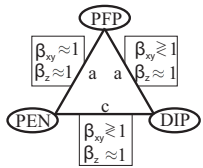}
  \caption{Length ratios $\beta$ of the molecules in the different mixed
    systems and the mixing behavior observed (\textit{a} or \textit{c}, see
    Fig.~\ref{fig:GrowthScen}).}
  \label{fig:Triangle}
\end{figure}

We account for the anisotropic ordering behavior observed for PEN:DIP mixtures
in the broader context of results on equimolar mixtures of
PEN:PFP~\cite{Hinderhofer2011J.Chem.Phys._134,Broch_2011_PRB_83_245307} and
PFP:DIP ~\cite{Reinhardt_2012_JPCC_116_10917} (see Fig.~\ref{fig:Triangle}),
considering all systems as mixtures of rod-like molecules and discussing the
interplay of the three contributions to $\Delta F$, which is influenced by
sterical properties and intermolecular interactions, defined by the chemical
composition of the two compounds.

PFP and PEN are sterically highly compatible, i.e. $\beta_{xy}\approx \beta_z
\approx 1$, leading to a low strain energy $E^s_{\alpha}$ in both directions
$\alpha=[xy,z]$.  The different charge distribution on PEN and PFP, induced by
the perfluorination, is expected to give rise to an attractive interaction,
i.e. $\chi_{xy}<0$, and results in the formation of a strongly coupled
molecular complex phase (with PEN:PFP 1:1) upon
mixing~\cite{Hinderhofer2011J.Chem.Phys._134,Broch_2011_PRB_83_245307}.
 
In blends of PFP:DIP, the sterical compatibility is lower compared to PEN:PFP,
since PFP and DIP have a slightly different shape (Fig.~\ref{fig:XRR}a), which
leads to $\beta_{xy} \gtrless 1$ while $\beta_z \approx 1$. The intermolecular
interaction between PFP and DIP, though, is expected to be comparable to that
between PFP and PEN, i.e.\ $\chi_{xy}<0$, due to the presence of fluorine in
the system and consequently strong electrostatic interaction between the
molecules. The resulting preference for A-B pairing, seems to outweigh the
increase in entropy by statistial mixing and the in-plane sterical
incompatibility, as for PFP:DIP blends the formation of a molecular complex
phase was found recently~\cite{Reinhardt_2012_JPCC_116_10917}.

Importantly, even though PEN:DIP blends exhibit similar sterical
characteristics compared to PFP:DIP, i.e. $\beta_{xy}\gtrless 1$ and $\beta_z
\approx 1$, the ordering behavior is completely different. In both systems a
good out-of-plane order is maintained due to $\beta_z \approx 1$. However,
while a mixed crystal phase was found for equimolar mixed PFP:DIP films,
equimolar PEN:DIP mixtures exhibit vanishing in-plane order and statistical
mixing of PEN and DIP.  This experimental observation is consistent with a
$\chi_{xy} \approx 0$, which can be rationalized by the chemical differences
of the PFP:DIP and PEN:DIP-blends. In particular, the presence of fluorine in
PFP and the associated strong quadrupole-moment are of importance, as the
quadrupole-quadrupole interaction is expected to be a key ingredient in
mixtures forming new crystal structures~\cite{Patrick1960Nature_187,
  Meyer2003ACIE_42}.  Due to the ``chemical similarity'' of PEN and DIP and
the corresponding lack of significant and specific A-B interactions, which
results in $\chi_{xy}\approx0$, only the entropy and strain terms contribute
to the free energy $\Delta F$. Consequently, the preference for entropy
dominated statistical mixing competes with the increase of $E^{s}_{xy}$ upon
mixing due to $\beta_{xy}\gtrless 1$.  As the entropy term seems to outweigh
the strain term, statistical mixing in the in-plane direction takes
place. This random occupation of nearest neighbor sites in combination with
the in-plane sterical incompatibility ($\beta_{xy}\gtrless 1$), prevents the
formation of a periodic structure in the in-plane direction and results in the
observed vanishing of in-plane order.

We note that our model also covers the case of phase separation in mixed
systems, as it is observed for mixtures of
PEN:$\mathrm{C_{60}}$~\cite{Salzmann_2008_JournalofAppliedPhysics_104} and
DIP:$\mathrm{C_{60}}$~\cite{Wagner2010Adv.Funct.Mater._20}. Here, the strain
energies E$^{s}_{xy}$ and E$^{s}_z$ resulting from the huge differences in
molecular shape dominate over the entropy terms and lead to phase
separation. With similar arguments this behavior can also be expected for
mixed films of PFP:$\mathrm{C_{60}}$.

%%%%%%%%%%%%%%%%%%%%%%%%%%%%%%%%%%%%%%%%%%%%%%%%%%%%%%%%%%%%%%%%%%
%%%%%%%%%%%%%
%%%%%%%%%%%%%           CONCLUSION          %%%%%%%%%%%%%%%%%%%%%%
%%%%%%%%%%%%%
%%%%%%%%%%%%%%%%%%%%%%%%%%%%%%%%%%%%%%%%%%%%%%%%%%%%%%%%%%%%%%%%%%
To conclude, we investigated the ordering behavior of PEN and DIP mixed
films. In the framework of sterical compatibility and the interaction
parameter $\chi$ we compared our results to data reported for blended films of
PEN:PFP~\cite{Hinderhofer2011J.Chem.Phys._134} and
PFP:DIP~\cite{Reinhardt_2012_JPCC_116_10917}. When mixing PEN and DIP, two
compounds, which are crystalline as pure materials as well as in binary blends
with PFP, we observed the reported effect of break-down of in-plane order upon
mixing. Equimolar PEN:DIP mixtures exhibit an anisotropical ordering behavior,
which is liquid-crystal like and comparable to a ``frozen'' smectic C phase,
with a well ordered out-of-plane structure but no detectable in-plane order.

Support from the DFG and the ESRF is gratefully acknowledged. K. Broch was
supported by the Studienstiftung des Deutschen Volkes. We thank K. Mecke,
H. L\"owen, H. Stark and C. Tschierske for helpful discussions.

%%%%%%%%%%%%%%%%%%%%%%%%%%%%%%%%%%%%%%%%%%%%%%%%%%%%%%%%%%%%%%%%%%
%%%%%%%%%%%%%
%%%%%%%%%%%%%           BIBLIOGRPHY        %%%%%%%%%%%%%%%%%%%%%%%
%%%%%%%%%%%%%
%%%%%%%%%%%%%%%%%%%%%%%%%%%%%%%%%%%%%%%%%%%%%%%%%%%%%%%%%%%%%%%%%%


\begin{thebibliography}{21}%
\expandafter\ifx\csname natexlab\endcsname\relax\def\natexlab#1{#1}\fi
\expandafter\ifx\csname bibnamefont\endcsname\relax
  \def\bibnamefont#1{#1}\fi
\expandafter\ifx\csname bibfnamefont\endcsname\relax
  \def\bibfnamefont#1{#1}\fi
\expandafter\ifx\csname citenamefont\endcsname\relax
  \def\citenamefont#1{#1}\fi
\expandafter\ifx\csname url\endcsname\relax
  \def\url#1{\texttt{#1}}\fi
\expandafter\ifx\csname urlprefix\endcsname\relax\def\urlprefix{URL }\fi
\providecommand{\bibinfo}[2]{#2}
\providecommand{\eprint}[2][]{\url{#2}}

\bibitem[{\citenamefont{Forrest}(2004)}]{Forrest2004N_428}
\bibinfo{author}{\bibfnamefont{S.~R.} \bibnamefont{Forrest}},
  \bibinfo{journal}{Nature} \textbf{\bibinfo{volume}{428}},
  \bibinfo{pages}{911} (\bibinfo{year}{2004}).

\bibitem[{\citenamefont{Br\"utting}(2005)}]{Bruetting_2005}
\bibinfo{editor}{\bibfnamefont{W.}~\bibnamefont{Br\"utting}}, ed.,
  \emph{\bibinfo{title}{Physics of Organic Semiconductors}}
  (\bibinfo{publisher}{Wiley-VCH, Weinheim}, \bibinfo{year}{2005}).

\bibitem[{\citenamefont{Brabec et~al.}(2001)\citenamefont{Brabec, Sariciftci,
  and Hummelen}}]{Brabec2001Afm_11}
\bibinfo{author}{\bibfnamefont{C.~J.} \bibnamefont{Brabec}},
  \bibinfo{author}{\bibfnamefont{N.~S.} \bibnamefont{Sariciftci}},
  \bibnamefont{and} \bibinfo{author}{\bibfnamefont{J.~C.}
  \bibnamefont{Hummelen}}, \bibinfo{journal}{Adv. Funct. Mater.}
  \textbf{\bibinfo{volume}{11}}, \bibinfo{pages}{15} (\bibinfo{year}{2001}).

\bibitem[{\citenamefont{Opitz et~al.}(2010)\citenamefont{Opitz, Wagner,
  Br\"utting, Salzmann, Koch, Manara, Pflaum, Hinderhofer, and
  Schreiber}}]{opitz_ieee10}
\bibinfo{author}{\bibfnamefont{A.}~\bibnamefont{Opitz}},
  \bibinfo{author}{\bibfnamefont{J.}~\bibnamefont{Wagner}},
  \bibinfo{author}{\bibfnamefont{W.}~\bibnamefont{Br\"utting}},
  \bibinfo{author}{\bibfnamefont{I.}~\bibnamefont{Salzmann}},
  \bibinfo{author}{\bibfnamefont{N.}~\bibnamefont{Koch}},
  \bibinfo{author}{\bibfnamefont{J.}~\bibnamefont{Manara}},
  \bibinfo{author}{\bibfnamefont{J.}~\bibnamefont{Pflaum}},
  \bibinfo{author}{\bibfnamefont{A.}~\bibnamefont{Hinderhofer}},
  \bibnamefont{and}
  \bibinfo{author}{\bibfnamefont{F.}~\bibnamefont{Schreiber}},
  \bibinfo{journal}{IEEE J. Sel. Top. Quant.} \textbf{\bibinfo{volume}{99}},
  \bibinfo{pages}{1} (\bibinfo{year}{2010}).
  %\urlprefix\url{http://dx.doi.org/10.1109/JSTQE.2010.2048096}.

\bibitem[{\citenamefont{Vogel et~al.}(2010)\citenamefont{Vogel, Salzmann, Duhm,
  Oehzelt, Rabe, and Koch}}]{Vogel_2010_JournalOfMaterialsChemistry_20}
\bibinfo{author}{\bibfnamefont{J.~O.} \bibnamefont{Vogel}},
  \bibinfo{author}{\bibfnamefont{I.}~\bibnamefont{Salzmann}},
  \bibinfo{author}{\bibfnamefont{S.}~\bibnamefont{Duhm}},
  \bibinfo{author}{\bibfnamefont{M.}~\bibnamefont{Oehzelt}},
  \bibinfo{author}{\bibfnamefont{J.~P.} \bibnamefont{Rabe}}, \bibnamefont{and}
  \bibinfo{author}{\bibfnamefont{N.}~\bibnamefont{Koch}}, \bibinfo{journal}{J.
  Mater. Chem.} \textbf{\bibinfo{volume}{20}}, \bibinfo{pages}{4055}
  (\bibinfo{year}{2010}).

\bibitem[{\citenamefont{Hinderhofer and
  Schreiber}(2012)}]{Hinderhofer_C_inprint}
\bibinfo{author}{\bibfnamefont{A.}~\bibnamefont{Hinderhofer}} \bibnamefont{and}
  \bibinfo{author}{\bibfnamefont{F.}~\bibnamefont{Schreiber}},
  \bibinfo{journal}{ChemPhysChem} \textbf{\bibinfo{volume}{13}},
  \bibinfo{pages}{628} (\bibinfo{year}{2012}).
  %\urlprefix\url{http://dx.doi.org/doi:10.1002/cphc.201100737}.

\bibitem[{\citenamefont{Kitaigorodsky}(1984)}]{Kitaigorodsky1984}
\bibinfo{author}{\bibfnamefont{A.}~\bibnamefont{Kitaigorodsky}},
  \emph{\bibinfo{title}{Mixed Crystals}} (\bibinfo{publisher}{Springer, Berlin,
  Heidelberg}, \bibinfo{year}{1984}).

\bibitem[{\citenamefont{Hinderhofer et~al.}(2011)\citenamefont{Hinderhofer,
  Frank, Hosokai, Resta, Gerlach, and
  Schreiber}}]{Hinderhofer2011J.Chem.Phys._134}
\bibinfo{author}{\bibfnamefont{A.}~\bibnamefont{Hinderhofer}},
  \bibinfo{author}{\bibfnamefont{C.}~\bibnamefont{Frank}},
  \bibinfo{author}{\bibfnamefont{T.}~\bibnamefont{Hosokai}},
  \bibinfo{author}{\bibfnamefont{A.}~\bibnamefont{Resta}},
  \bibinfo{author}{\bibfnamefont{A.}~\bibnamefont{Gerlach}}, \bibnamefont{and}
  \bibinfo{author}{\bibfnamefont{F.}~\bibnamefont{Schreiber}},
  \bibinfo{journal}{J. Chem. Phys.} \textbf{\bibinfo{volume}{134}},
  \bibinfo{pages}{104702} (\bibinfo{year}{2011}).
  %\urlprefix\url{http://link.aip.org/link/?JCP/134/104702/1}.

\bibitem[{\citenamefont{Reinhardt et~al.}(2012)\citenamefont{Reinhardt,
  Hinderhofer, Broch, Heinemeyer, Kowarik, Gerlach, and
  Schreiber}}]{Reinhardt_2012_JPCC_116_10917}
\bibinfo{author}{\bibfnamefont{J.}~\bibnamefont{Reinhardt}},
  \bibinfo{author}{\bibfnamefont{A.}~\bibnamefont{Hinderhofer}},
  \bibinfo{author}{\bibfnamefont{K.}~\bibnamefont{Broch}},
  \bibinfo{author}{\bibfnamefont{U.}~\bibnamefont{Heinemeyer}},
  \bibinfo{author}{\bibfnamefont{S.}~\bibnamefont{Kowarik}},
  \bibinfo{author}{\bibfnamefont{A.}~\bibnamefont{Gerlach}}, \bibnamefont{and}
  \bibinfo{author}{\bibfnamefont{F.}~\bibnamefont{Schreiber}},
  \bibinfo{journal}{J. Phys. Chem. C} \textbf{\bibinfo{volume}{116}},
  \bibinfo{pages}{10917} (\bibinfo{year}{2012}).

\bibitem[{\citenamefont{de~Oteyza et~al.}(2008)\citenamefont{de~Oteyza,
  Barrena, Sellner, Oss\'o, and Dosch}}]{Oteyza_2008_ThinSolidFilms_516}
\bibinfo{author}{\bibfnamefont{D.}~\bibnamefont{de~Oteyza}},
  \bibinfo{author}{\bibfnamefont{E.}~\bibnamefont{Barrena}},
  \bibinfo{author}{\bibfnamefont{S.}~\bibnamefont{Sellner}},
  \bibinfo{author}{\bibfnamefont{J.}~\bibnamefont{Oss\'o}}, \bibnamefont{and}
  \bibinfo{author}{\bibfnamefont{H.}~\bibnamefont{Dosch}},
  \bibinfo{journal}{Thin Solid Films} \textbf{\bibinfo{volume}{516}},
  \bibinfo{pages}{7525 } (\bibinfo{year}{2008}).%, ISSN
  \bibinfo{issn}{0040-6090}.
  %\urlprefix\url{http://www.sciencedirect.com/science/article/B6TW0-4SC78MV-1/%
%2/54b638fd629a3cb96d79709823be3d19}.

\bibitem[{\citenamefont{Opitz et~al.}(2009)\citenamefont{Opitz, Ecker, Wagner,
  Hinderhofer, Schreiber, Manara, Pflaum, and Br\"utting}}]{opitz_oe09}
\bibinfo{author}{\bibfnamefont{A.}~\bibnamefont{Opitz}},
  \bibinfo{author}{\bibfnamefont{B.}~\bibnamefont{Ecker}},
  \bibinfo{author}{\bibfnamefont{J.}~\bibnamefont{Wagner}},
  \bibinfo{author}{\bibfnamefont{A.}~\bibnamefont{Hinderhofer}},
  \bibinfo{author}{\bibfnamefont{F.}~\bibnamefont{Schreiber}},
  \bibinfo{author}{\bibfnamefont{J.}~\bibnamefont{Manara}},
  \bibinfo{author}{\bibfnamefont{J.}~\bibnamefont{Pflaum}}, \bibnamefont{and}
  \bibinfo{author}{\bibfnamefont{W.}~\bibnamefont{Br\"utting}},
  \bibinfo{journal}{Organic Electronics} \textbf{\bibinfo{volume}{10}},
  \bibinfo{pages}{1259} (\bibinfo{year}{2009}).
  %\urlprefix\url{http://dx.doi.org/doi:10.1016/j.orgel.2009.07.004}.

\bibitem[{\citenamefont{Chen et~al.}(2011)\citenamefont{Chen, Qi, Huang, Gao,
  and Wee}}]{Chen_2011_Adv.Funct.Mater._21}
\bibinfo{author}{\bibfnamefont{W.}~\bibnamefont{Chen}},
  \bibinfo{author}{\bibfnamefont{D.-C.} \bibnamefont{Qi}},
  \bibinfo{author}{\bibfnamefont{H.}~\bibnamefont{Huang}},
  \bibinfo{author}{\bibfnamefont{X.}~\bibnamefont{Gao}}, \bibnamefont{and}
  \bibinfo{author}{\bibfnamefont{A.~T.~S.} \bibnamefont{Wee}},
  \bibinfo{journal}{Adv. Funct. Mater.} \textbf{\bibinfo{volume}{21}},
  \bibinfo{pages}{410} (\bibinfo{year}{2011}).%, ISSN \bibinfo{issn}{1616-3028}.
  %\urlprefix\url{http://dx.doi.org/10.1002/adfm.201000902}.

\bibitem[{\citenamefont{Campione et~al.}(2009)\citenamefont{Campione, Raimondo,
  Moret, Campiglio, Fumagalli, and Sassella}}]{Campione2009Chem.Mater._21}
\bibinfo{author}{\bibfnamefont{M.}~\bibnamefont{Campione}},
  \bibinfo{author}{\bibfnamefont{L.}~\bibnamefont{Raimondo}},
  \bibinfo{author}{\bibfnamefont{M.}~\bibnamefont{Moret}},
  \bibinfo{author}{\bibfnamefont{P.}~\bibnamefont{Campiglio}},
  \bibinfo{author}{\bibfnamefont{E.}~\bibnamefont{Fumagalli}},
  \bibnamefont{and} \bibinfo{author}{\bibfnamefont{A.}~\bibnamefont{Sassella}},
  \bibinfo{journal}{Chem. Mater.} \textbf{\bibinfo{volume}{21}},
  \bibinfo{pages}{4859} (\bibinfo{year}{2009}).
  %\urlprefix\url{http://pubs.acs.org/doi/abs/10.1021/cm901463u}.

\bibitem[{\citenamefont{Salzmann et~al.}(2008)\citenamefont{Salzmann, Duhm,
  Opitz, Johnson, Rabe, and Koch}}]{Salzmann_2008_JournalofAppliedPhysics_104}
\bibinfo{author}{\bibfnamefont{I.}~\bibnamefont{Salzmann}},
  \bibinfo{author}{\bibfnamefont{S.}~\bibnamefont{Duhm}},
  \bibinfo{author}{\bibfnamefont{R.}~\bibnamefont{Opitz}},
  \bibinfo{author}{\bibfnamefont{R.~L.} \bibnamefont{Johnson}},
  \bibinfo{author}{\bibfnamefont{J.~P.} \bibnamefont{Rabe}}, \bibnamefont{and}
  \bibinfo{author}{\bibfnamefont{N.}~\bibnamefont{Koch}}, \bibinfo{journal}{J.
  Appl. Phys.} \textbf{\bibinfo{volume}{104}}, \bibinfo{pages}{114518}
  (\bibinfo{year}{2008}).
  %\urlprefix\url{http://link.aip.org/link/?JAP/104/114518/1}.

\bibitem[{\citenamefont{Broch et~al.}(2011)\citenamefont{Broch, Heinemeyer,
  Hinderhofer, Anger, Scholz, Gerlach, and
  Schreiber}}]{Broch_2011_PRB_83_245307}
\bibinfo{author}{\bibfnamefont{K.}~\bibnamefont{Broch}},
  \bibinfo{author}{\bibfnamefont{U.}~\bibnamefont{Heinemeyer}},
  \bibinfo{author}{\bibfnamefont{A.}~\bibnamefont{Hinderhofer}},
  \bibinfo{author}{\bibfnamefont{F.}~\bibnamefont{Anger}},
  \bibinfo{author}{\bibfnamefont{R.}~\bibnamefont{Scholz}},
  \bibinfo{author}{\bibfnamefont{A.}~\bibnamefont{Gerlach}}, \bibnamefont{and}
  \bibinfo{author}{\bibfnamefont{F.}~\bibnamefont{Schreiber}},
  \bibinfo{journal}{Phys. Rev. B} \textbf{\bibinfo{volume}{83}},
  \bibinfo{pages}{245307} (\bibinfo{year}{2011}).

\bibitem[{\citenamefont{Warren}(1991)}]{Warren_1991_DoverPubnIncDover}
\bibinfo{author}{\bibfnamefont{B.}~\bibnamefont{Warren}},
  \emph{\bibinfo{title}{X-ray diffraction}} (\bibinfo{publisher}{Dover Pubn
  Inc., Dover}, \bibinfo{year}{1991}).

\bibitem[{\citenamefont{de~Gennes and Prost}(1993)}]{Gennes_1993}
\bibinfo{author}{\bibfnamefont{P.}~\bibnamefont{de~Gennes}} \bibnamefont{and}
  \bibinfo{author}{\bibfnamefont{J.}~\bibnamefont{Prost}},
  \emph{\bibinfo{title}{The Physics of Liquid Crystals}}
  (\bibinfo{publisher}{Clarendon Press Oxford}, \bibinfo{year}{1993}).

\bibitem[{\citenamefont{Huang et~al.}(2011)\citenamefont{Huang, McCreary, Garg,
  and Kyu}}]{Huang_2011_JournalofChemicalPhysics_134_124508}
\bibinfo{author}{\bibfnamefont{T.-M.} \bibnamefont{Huang}},
  \bibinfo{author}{\bibfnamefont{K.}~\bibnamefont{McCreary}},
  \bibinfo{author}{\bibfnamefont{S.}~\bibnamefont{Garg}}, \bibnamefont{and}
  \bibinfo{author}{\bibfnamefont{T.}~\bibnamefont{Kyu}}, \bibinfo{journal}{J.
  Chem. Phys.} \textbf{\bibinfo{volume}{134}}, \bibinfo{pages}{124508}
  (\bibinfo{year}{2011}).

\bibitem[{\citenamefont{Rauch et~al.}(2002)\citenamefont{Rauch, Garg, and
  Jacobs}}]{Rauch_2002_JCP_116_2213}
\bibinfo{author}{\bibfnamefont{A.~C.} \bibnamefont{Rauch}},
  \bibinfo{author}{\bibfnamefont{S.}~\bibnamefont{Garg}}, \bibnamefont{and}
  \bibinfo{author}{\bibfnamefont{D.~T.} \bibnamefont{Jacobs}},
  \bibinfo{journal}{J. Chem. Phys.} \textbf{\bibinfo{volume}{116}},
  \bibinfo{pages}{2213} (\bibinfo{year}{2002}).

\bibitem[{\citenamefont{Kapernaum et~al.}(2010)\citenamefont{Kapernaum,
  Hartley, Roberts, Schoerg, Krueerke, Lemieux, and
  Giesselmann}}]{Kapernaum_2010_C_11_2099}
\bibinfo{author}{\bibfnamefont{N.}~\bibnamefont{Kapernaum}},
  \bibinfo{author}{\bibfnamefont{C.~S.} \bibnamefont{Hartley}},
  \bibinfo{author}{\bibfnamefont{J.~C.} \bibnamefont{Roberts}},
  \bibinfo{author}{\bibfnamefont{F.}~\bibnamefont{Schoerg}},
  \bibinfo{author}{\bibfnamefont{D.}~\bibnamefont{Krueerke}},
  \bibinfo{author}{\bibfnamefont{R.~P.} \bibnamefont{Lemieux}},
  \bibnamefont{and}
  \bibinfo{author}{\bibfnamefont{F.}~\bibnamefont{Giesselmann}},
  \bibinfo{journal}{ChemPhysChem} \textbf{\bibinfo{volume}{11}},
  \bibinfo{pages}{2099} (\bibinfo{year}{2010}).
  
\bibitem[{\citenamefont{C. R. Patrick and G. S. Prosser}(1960)\citenamefont{Patrick, Prosser}}]{Patrick1960Nature_187}
\bibinfo{author}{\bibfnamefont{C. R.}~\bibnamefont{Patrick}},
\bibinfo{author}{\bibfnamefont{G. S.}~\bibnamefont{Prosser}},
  \bibinfo{journal}{Nature} \textbf{\bibinfo{volume}{187}},
  \bibinfo{pages}{1021} (\bibinfo{year}{1960}).

\bibitem[{\citenamefont{E. A. Meyer et~al.}(1960)\citenamefont{Meyer, Castellano, Diederich}}]{Meyer2003ACIE_42}
\bibinfo{author}{\bibfnamefont{E. A.}~\bibnamefont{Meyer}},
\bibinfo{author}{\bibfnamefont{R. K.}~\bibnamefont{Castellano}},
\bibinfo{author}{\bibfnamefont{F.}~\bibnamefont{Diederich}},
  \bibinfo{journal}{Angew. Chem. Int. Ed.} \textbf{\bibinfo{volume}{42}},
  \bibinfo{pages}{1210} (\bibinfo{year}{2003}).

\bibitem[{\citenamefont{Wagner et~al.}(2010)\citenamefont{Wagner, Gruber,
  Hinderhofer, Wilke, Br{\"o}ker, Frisch, Amsalem, Vollmer, Opitz, Koch
  et~al.}}]{Wagner2010Adv.Funct.Mater._20}
\bibinfo{author}{\bibfnamefont{J.}~\bibnamefont{Wagner}},
  \bibinfo{author}{\bibfnamefont{M.}~\bibnamefont{Gruber}},
  \bibinfo{author}{\bibfnamefont{A.}~\bibnamefont{Hinderhofer}},
  \bibinfo{author}{\bibfnamefont{A.}~\bibnamefont{Wilke}},
  \bibinfo{author}{\bibfnamefont{B.}~\bibnamefont{Br{\"o}ker}},
  \bibinfo{author}{\bibfnamefont{J.}~\bibnamefont{Frisch}},
  \bibinfo{author}{\bibfnamefont{P.}~\bibnamefont{Amsalem}},
  \bibinfo{author}{\bibfnamefont{A.}~\bibnamefont{Vollmer}},
  \bibinfo{author}{\bibfnamefont{A.}~\bibnamefont{Opitz}},
  \bibinfo{author}{\bibfnamefont{N.}~\bibnamefont{Koch}}, \bibnamefont{et~al.},
  \bibinfo{journal}{Adv. Funct. Mater.} \textbf{\bibinfo{volume}{20}},
  \bibinfo{pages}{4295} (\bibinfo{year}{2010}).
  %\urlprefix\url{http://dx.doi.org/10.1002/adfm.201001028}.
  

\end{thebibliography}
\end{document}